\documentclass[preprintnumbers,superscriptaddress,amsmath,amssymb,prd,preprint,nofootinbib]{revtex4-2}
\usepackage{graphicx}
\usepackage{amsfonts}
\usepackage[colorlinks,linkcolor=red,anchorcolor=blue,citecolor=green]{hyperref}
\usepackage{subfigure}
\usepackage{float}
\usepackage{enumerate}
\newcommand\diff{\mathrm{d}}

\begin{document}
\title{
    Viscous effect in the late time evolution of phantom universe
}
\author{Jing Yang}
\author{Rui-Hui Lin}
\email[]{linrh@shnu.edu.cn}
\author{Chao-Jun Feng}
\email[]{fengcj@shnu.edu.cn}
\author{Xiang-Hua Zhai}
\email[]{zhaixh@shnu.edu.cn}
\affiliation{Division of Mathematics and Theoretical Physics, Shanghai Normal University, 100 Guilin Road, Shanghai 200234, China}

\begin{abstract}
    We investigate the cosmological implications of a phantom dark energy model with bulk viscosity.
    We explore this model as a possible way to resolve the big rip singularity problem that plagues the phantom models.
    We use the latest type Ia supernova and Hubble parameter data to constrain the model parameters
    and find that the data favor a significant bulk viscosity over a non-constant potential term for the phantom field.
    We perform a dynamical analysis of the model and
    show that the only stable and physical attractor corresponds to a phantom-dominated era with a total equation of state that can be greater than $-1$ due to the viscosity.
    We also study the general effect of viscosity on the phantom field and the late time evolution of the universe.
    We apply the statefinder diagnostic to the model and find that it approaches a nearby fixed point asymptotically,
    indicating that the universe can escape the big rip singularity with the presence of bulk viscosity.
    We conclude that bulk viscosity can play an important role in affecting the late-time behavior
    as well as alleviating the singularity problem of the phantom universe.
\end{abstract}
\maketitle
\newpage
\section{Introduction}
\label{intro}
Cosmological observations in recent decades have revealed more details about the evolutional history of the universe.
One of the major discoveries is the late time acceleration of cosmic expansion,
which can be generally understood from at least two different perspectives.
Either Einstein’s theory of gravity is incomplete and calls for modifications
(for recent reviews, see, e.g., Refs. \cite{Capozziello:2011et,Nojiri:2017ncd}),
or there is an unknown form of energy dubbed dark energy (DE) that exhibits repulsive behavior.
The simplest model of DE is the cosmological constant $\Lambda$ model with cold dark matter ($\Lambda$CDM),
which assumes that DE consists of a constant energy density that fills the space uniformly
and has a constant equation of state (EoS) $w_\text{DE}=-1$,
where $w_\text{DE}$ is the ratio of pressure to energy density of the DE content.

However, the cosmological constant has no clear physical origin and faces several theoretical challenges (see, e.g., Ref. \cite{Weinberg2000}).
Therefore, many alternative DE models have been proposed and explored (see, e.g., Refs. \cite{Bahamonde:2017ize,Pan-STARRS1:2017jku,Ishak:2018his}).
These models can be roughly divided into two categories:
the quintessence model with $w_\text{DE}>-1$\cite{Peebles1988,Ratra1988} and the phantom model with $w_\text{DE}<-1$\cite{Caldwell:1999ew,Carroll2003},
which have different implications for the ultimate fate of the universe.
Quintessence models generally predict that the universe will undergo an eternal expansion,
while phantom models usually indicate that the universe will enter a super-accelerated expansion phase
and end in a finite time with a cosmic singularity known as big rip, where all structures will be torn apart.
Recent observations seem to favor the phantom model over the quintessence model
\cite{Planck:2018vyg,BeltranJimenez:2016fuy,Caldwell:1999ew,Caldwell2003,Sahni:2014ooa}.
Nonetheless, such discussions about $w_\text{DE}$ rest upon the assumption that it is simply given by $p_\text{DE}/\rho_\text{DE}$.
This representation is purely phenomenological and lacks underlying physics.
To provide a more comprehensive understanding of the nature of DE including its temporal evolution or perturbation,
a more fundamental framework should be considered.
One approach involves introducing a dynamic scalar field as part of either the quintessence or phantom models,
depending on the specific properties of its potential and kinetic terms\cite{Caldwell:1999ew}.
In particular, the phantom scalar field model has attracted significant interests within the literature
(see, e.g., Refs. \cite{Bahamonde:2017ize,Caldwell2003,Sahni:2004ai} and the references therein).

The studies of phantom models typically lead to discussions about the big rip singularity
\cite{Sami:2003xv,Guo:2004xx,Guo:2004vg,Nojiri:2005sx,Chang:2006dj,Curbelo:2005dh,Setare:2007eq,Fu:2008gh,deHaro:2012pu,Amirhashchi:2013sua}.
To address this issue, researchers have introduced modifications to DE based on quantum effects\cite{Sami:2006wj},
geometric effects derived from modified gravities\cite{Koivisto:2006xf,Nojiri:2008fk},
or interactions between cosmic content\cite{Huey:2004qv,Wei:2007ut}.
One way to introduce interactions involves considering the viscosity of each component.
This approach accounts for the dissipative properties of real fluids.
In particular, bulk viscosity is the most relevant for cosmology since we assume a homogenous and isotropic universe.
This can be incorporated into the standard cosmological scheme
by redefining the effective pressure $p_\text{eff}=p-\Pi$ with a viscous pressure term $\Pi$ to restore thermal equilibrium\cite{OKUMURA2003207,Disconzi:2014oda,Brevik:2017msy}.

Invoking viscosity in cosmology is proven to be useful to resolve or soften the cosmic singularity problem in different models
\cite{Cataldo:2005qh,PhysRevD.84.103508,Sebastiani2011,Meng2012,Brevik2015,Boko2020,Singh2020,Cruz2022}.
Following this approach, it is shown that the singularity problem can be alleviated in the anisotropic phantom universe with viscosity\cite{Amirhashchi:2013sua}.
Moreover, an interacting phantom DE with dark matter induced by the viscous approach is shown to be able to avoid the big rip singularity\cite{Guo:2004xx,Guo:2004vg}.

In this work, we will study viscous phantom scalar field DE model and explore its late time behavior.
The paper is structured as follows.
In Sec. \ref{review}, we review the phantom scalar field model of DE and introduce the viscous cosmology framework.
Section \ref{fitting} contains the late time observation fit to constrain the model parameters.
Dynamical system analysis and statefinder diagnostic of the model are given in Secs. \ref{dynamics} and \ref{statefinder}, respectively.
And we conclude our study in the last section.

\section{Viscous phantom scalar field DE model}
\label{review}
The action for a phantom field minimally coupled to gravity is given by
\begin{equation}
    \mathcal S=\int \diff^4x\sqrt{-g}\left[-(\partial \phi)^2+V(\phi) \right]\,,
\end{equation}
where $V(\phi)$ is the potential of the phantom field $\phi$.
The energy density and pressure of the phantom field are\cite{Caldwell:1999ew}
\begin{equation}\label{denphi}
    \begin{split}
        \rho_\phi=&-\frac{1}{2}\dot{\phi}^{2}+V(\phi)\,,\\
        p_\phi=&-\frac{1}{2}\dot{\phi}^{2}-V(\phi)\,.
    \end{split}
\end{equation}
We assume a spatially flat Friedmann-Lema\^itre-Robertson-Walker metric for the homogeneous and isotropic universe, given by
\begin{equation}
    \diff s^2=\diff t^2-a(t)^2\left[\diff r^2+r^2\left(\diff\theta^2+\sin^2\theta\diff\psi^2\right)\right],
\end{equation}
where $a(t)$ is the scale factor.
Then, the Friedmann equations read
\begin{equation}\label{Fri}
    \begin{split}
        H^{2}&=\frac{1}{3}\left( \rho_m+\rho_\phi \right)\,,\\
        \dot{H}&=-\frac{1}{2}\left( \rho_m+\rho_\phi+p_\phi \right)\,,
    \end{split}
\end{equation}
where $\rho_m$ and $p_m$ are the energy density and pressure of dust matter, respectively,
$ H=\frac{\dot{a}}{a}$ is the Hubble parameter, and the dot represents the derivative with respect to the cosmic time $t$.

As mentioned in the Introduction,
the bulk viscosity dissipation in the cosmic phantom field fluid can be represented by a pressure term $-\Pi$ added to $p_\phi$.
For bulk viscosity related to the cosmic expansion, we assume that $\Pi\propto H$, which implies that the effective pressure of the phantom field is
\begin{equation}
    p_\text{eff}=p_\phi-3\xi_\phi H,
\end{equation}
where $\xi_\phi$ is the viscosity coefficient.
Then, the evolutionary equations for dust matter and the phantom field are given by
\begin{equation}\label{eveq}
    \begin{split}
        \dot{\rho}_m+3H\rho_m&=0\,,\\
        \dot{\rho}_\phi+3H(\rho_\phi+p_\phi-3H\xi_\phi)&=0\,.
    \end{split}
\end{equation}
Using Eq. \eqref{denphi}, we obtain the equation for the phantom field
\begin{equation}\label{eqphi}
    \ddot{\phi}+3H\dot{\phi}-V'(\phi)+\frac{9\xi_\phi H^2}{\dot{\phi}}=0\,,
\end{equation}
where the prime denotes the derivative of the potential $V$ with respect to the field $\phi$.

Generally, the viscosity of a fluid may depend on energy density, pressure, spacetime geometry and so on.
For simplicity, in the current work, we assume the viscosity is proportional to the changing rate of the phantom field, i.e.,
\begin{equation}
    \xi_\phi=\xi_0\dot{\phi}\,,
\end{equation}
where the viscosity coefficient $\xi_{0}$ is a constant parameter.
As for the potential of the phantom field, it is shown that an exponential form of potential can match or account for the acceleration of expansion
\cite{Copeland1998,Ng2001,Hao2003,Hao2004,Li2004,Li2005},
which is then adopted in the current work and is given by
\begin{equation}
    V(\phi)=V_0 \rm{e}^{-\alpha_{0} \phi}\,,
\end{equation}
where $\alpha_{0}$ and $V_0$ are constants.

\section{Observational constraints}
\label{fitting}
\begin{figure}
    \centering
    \includegraphics[width=0.8\linewidth]{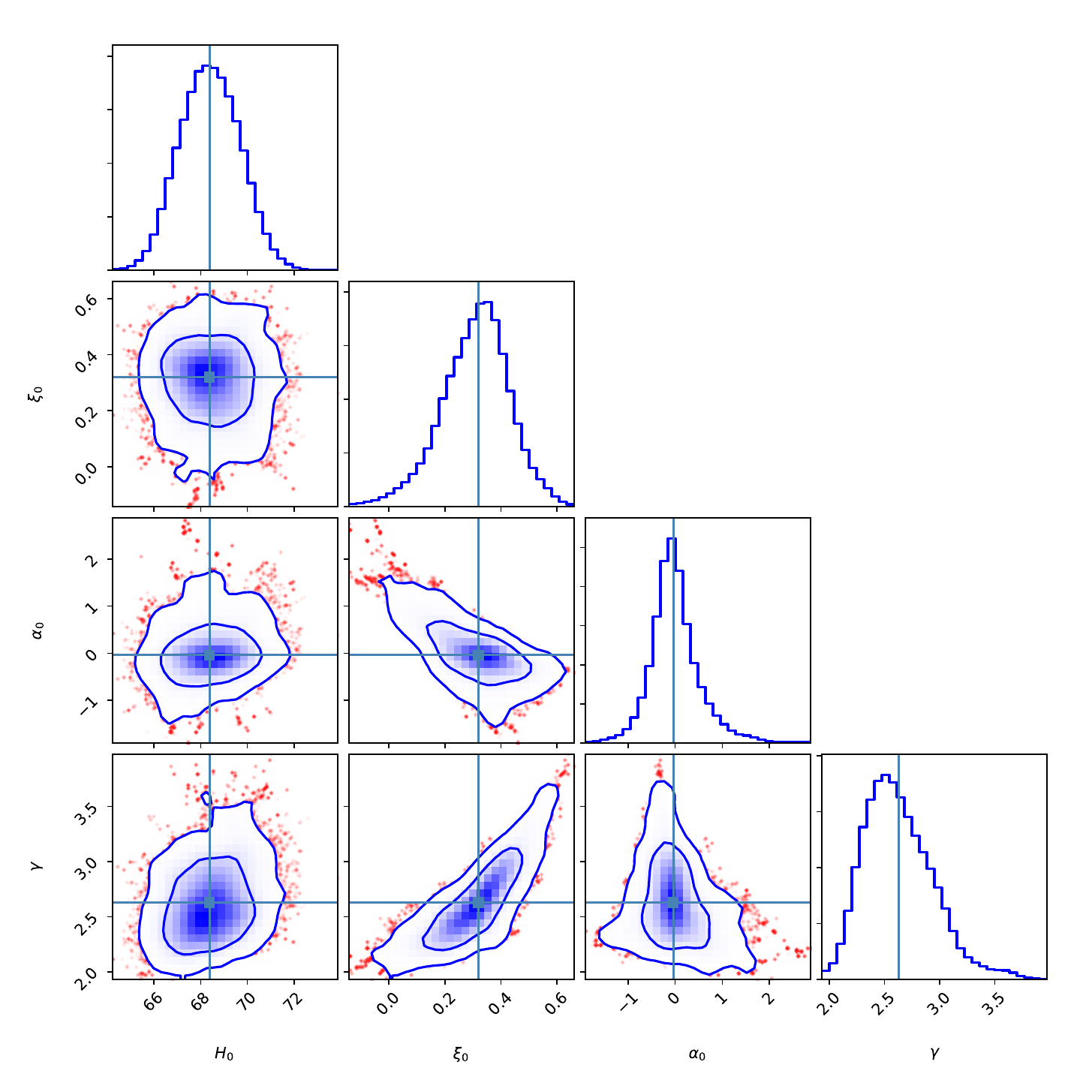}
    \caption{Constraints on the viscous phantom model parameters from 1$\sigma$ to 2$\sigma $ confidence level. Where $\gamma=V_{0}/H_{0}^{2}$.}
    \label{vphantom}
\end{figure}

We constrain the viscous phantom model using the late-time observational data sets based on Markov Chain Monte Carlo method.
We use the Pantheon compilation of Type Ia supernova (SNIa)\cite{Pan-STARRS1:2017jku} and Hubble parameter ($H(z)$) data points\cite{Sharov:2018yvz}
to fit the model parameters.
The SNIa data consist of 279 samples
from Sloan Digital Sky Survey (SDSS) and Supernova Legacy Survey (SNLS) with redshifts $0.03<z<0.68$,
and 1048 samples with redshifts $0.01<z<2.3$ including the Hubble Space Telescope (HST) samples and various low-$z$ samples.
The $H(z)$ data include 26 data points from Baryon Acoustic Oscillations and 31 data points from the differential age method.
For comparison, besides fitting the viscous phantom model under consideration (denoted as Model vP in the following),
we also perform the same fitting procedure for the phantom model without viscosity (denoted as Model P in the following).
The fitting results are summarized in Table \ref{fit}\footnote{For the detailed description of the fitting procedure, see Refs.\cite{Yang2022,Yang2023}}.
Figure \ref{vphantom} shows the constraints on the parameters of Model vP at 2$\sigma$ confidence level.
\begin{table}[!h]
    \centering
    \resizebox{!}{!}{
        \begin{tabular}{ccc}
            \hline
            \hline
            Parameter             & Model P                   & Model vP                   \\
            \hline
            $\Omega_m$            & $0.331^{+0.026}_{-0.025}$ & $0.322^{+0.003}_{ -0.003}$ \\
            $H_{0}$               & $68.31^{+1.16}_{-1.14}$   & $68.31^{+1.23}_{ -1.20}$   \\
            $V_{0}/H^{2}_{0}$     & $2.184^{+0.102}_{-0.087}$ & $2.612^{+0.347}_{-0.261}$  \\
            $\alpha_{0}$          & $1.262^{+0.479}_{-0.436}$ & $-0.04^{+0.355}_{-0.309}$  \\
            $\xi_{0}$             & $-$                       & $ 0.323^{+0.099}_{-0.106}$ \\
            \hline
            $\chi^{2}_{min}/dof.$ & $1053.93/1099$            & $1053.174/1098$            \\
            \hline
        \end{tabular}}
    \caption{The best fitting values of model parameters and $1\sigma$ confidence level for Model P and Model vP under Pantheon+H(z) data sets.}
    \label{fit}
\end{table}

One can see that for Model P, the phantom potential cannot be constant at $1\sigma$ confidence level
as the phantom field is the sole mechanism in operation for the late-time cosmic acceleration.
However, if both the viscosity and the phantom field with its potential are taken into consideration, as in Model vP,
the fitting result favors a significant viscosity and does not exclude the constant potential case with $\alpha_0=0$.

\section{Dynamical analysis}
\label{dynamics}
By using the dimensionless variables\cite{Copeland1998,Ng2001,Steinhardt1999}
\begin{equation}\label{xy}
    \begin{split}
        x=&\frac{\dot{\phi}}{\sqrt{6}H},\quad y=\frac{\sqrt{V(\phi)}}{\sqrt{3}H},\\
        \zeta=&\frac{\xi_\phi}{H}=\xi_0\frac{\dot{\phi}}{H}=\sqrt{6}\xi_0x,
    \end{split}
\end{equation}
the partial differential equations of the phantom scalar field can be recast into an autonomous system as
\begin{equation}\label{auto}
    \begin{split}
        \frac{\diff x}{\diff N}=&-3x-\frac{\sqrt{6}}{2}\alpha_{0} y^2+\frac{3}{2}x\bigg[1-x^2-y^2-\sqrt{6}\xi_0 x\bigg]-\frac{3\sqrt{6}}{2}\xi_0\,,\\
        \frac{\diff y}{\diff N}=&-\frac{\sqrt{6}}{2}\alpha_{0} xy+\frac{3}{2}y\bigg[1-x^2-y^2-\sqrt{6}\xi_0x \bigg]\,,
    \end{split}
\end{equation}
where $N=\ln a=-\ln (1+z)$.
The Friedmann equation can also be rewritten as a dimensionless constraint
\begin{equation}
    \Omega_m+\Omega_\phi=\Omega_m-x^2+y^2=1\,,
\end{equation}
where $\Omega_{i}=\frac{\rho_i}{3H^{2}}$ is the density parameter of the corresponding component.

The critical points of the above system and their existence conditions are listed in Table \ref{tab:m1}.
\begin{table}[!h]
    \centering
    \resizebox{!}{!}{
        \begin{tabular}{c|c|c|c}
            \hline
            \hline
            Critical points & $(x,y)$                                                                                              & Existence conditions                                                                                                     & $\Omega_{\phi}$                                  \\ \hline
            $A_{1}$         & $(\frac{\sqrt\frac{3}{2}}{\alpha_{0}}, \frac{\sqrt{ -\frac{3}{2}-3\alpha_{0} \xi_{0}}}{\alpha_{0}})$ & $\left(\xi_{0}<0 \& \alpha_{0}>-\frac{1}{2\xi_{0}}\right)$ or $\left(\xi_{0}>0 \& \alpha_{0}<-\frac{1}{2\xi_{0}}\right)$ & $-\frac{3(1+\alpha_{0}\xi_{0})}{\alpha_{0}^{2}}$ \\ \hline
            $A_{2}$         & $(-\sqrt{6}\xi_{0}, 0)$                                                                              & all $\xi_{0}$ and $\alpha_{0}\ne 0$                                                                                      & $-6\xi_{0}^{2}$                                  \\ \hline
            $A_{3}$         & $(-\frac{\alpha_{0}+3\xi_{0}}{\sqrt{6}}, \frac{\sqrt{6+(\alpha_{0}+3\xi_{0})^{2}}}{\sqrt{6}})$       & all $\xi_{0}$ and $\alpha_{0}\ne 0$                                                                                      & 1                                                \\
            \hline
            \hline
        \end{tabular}}
    \caption{The critical points and their existence conditions of autonomous system \eqref{auto}.}
    \label{tab:m1}
\end{table}
The stability of each critical point can be inferred from the eigenvalues
of the linearized perturbation matrix $\mathcal M$ of the autonomous given by
\begin{equation}\label{matrix}
    \mathcal{M}=\left(
    \begin{array}{cc}
            -\frac{3}{2}\big(1+3x^2+y^2+2\sqrt{6}\xi_{0}x\big)    & -3xy-\sqrt{6}\alpha_{0} y                                                         \\
            \\
            -3xy-\sqrt{6}\alpha_{0} y-\frac{3\sqrt{6}}{2}\xi_{0}y & -\frac{1}{2}\big(3+3x^2+9y^2+\sqrt{6}\alpha_{0} x+\frac{3\sqrt{6}}{2}\xi_0 x\big) \\
        \end{array}
    \right)\,.
\end{equation}

\textbullet\textbf{For critical point $A_{1}$},
the eigenvalues of the linearized perturbation matrix are
\begin{equation}
    \begin{split}
        \eta_{A_{1}}^{(1)}=&-\frac{3C_{1}+C_{2}}{4\alpha_{0}^{4}}\,,\\
        \eta _{A_{1}}^{(2)}=&-\frac{3C_{1}-C_{2}}{4\alpha_{0}^{4}}\,,
    \end{split}
\end{equation}
where
\begin{equation}
    \begin{split}
        C_{1}=&\alpha_{0}^{4}-3\alpha_{0}^{3}\xi_{0}\,,\\
        C_{2}=&\sqrt{-6\alpha_{0}^{6}\left[24+78\alpha_{0}\xi_{0}+16\alpha_{0}^{3}\xi_{0}-9\xi_{0}^{2}+\alpha_{0}^{2}(7+48\xi_{0}^{2})\right]}\,.
    \end{split}
\end{equation}
\begin{figure}
    \centering
    \includegraphics[width=.9\linewidth]{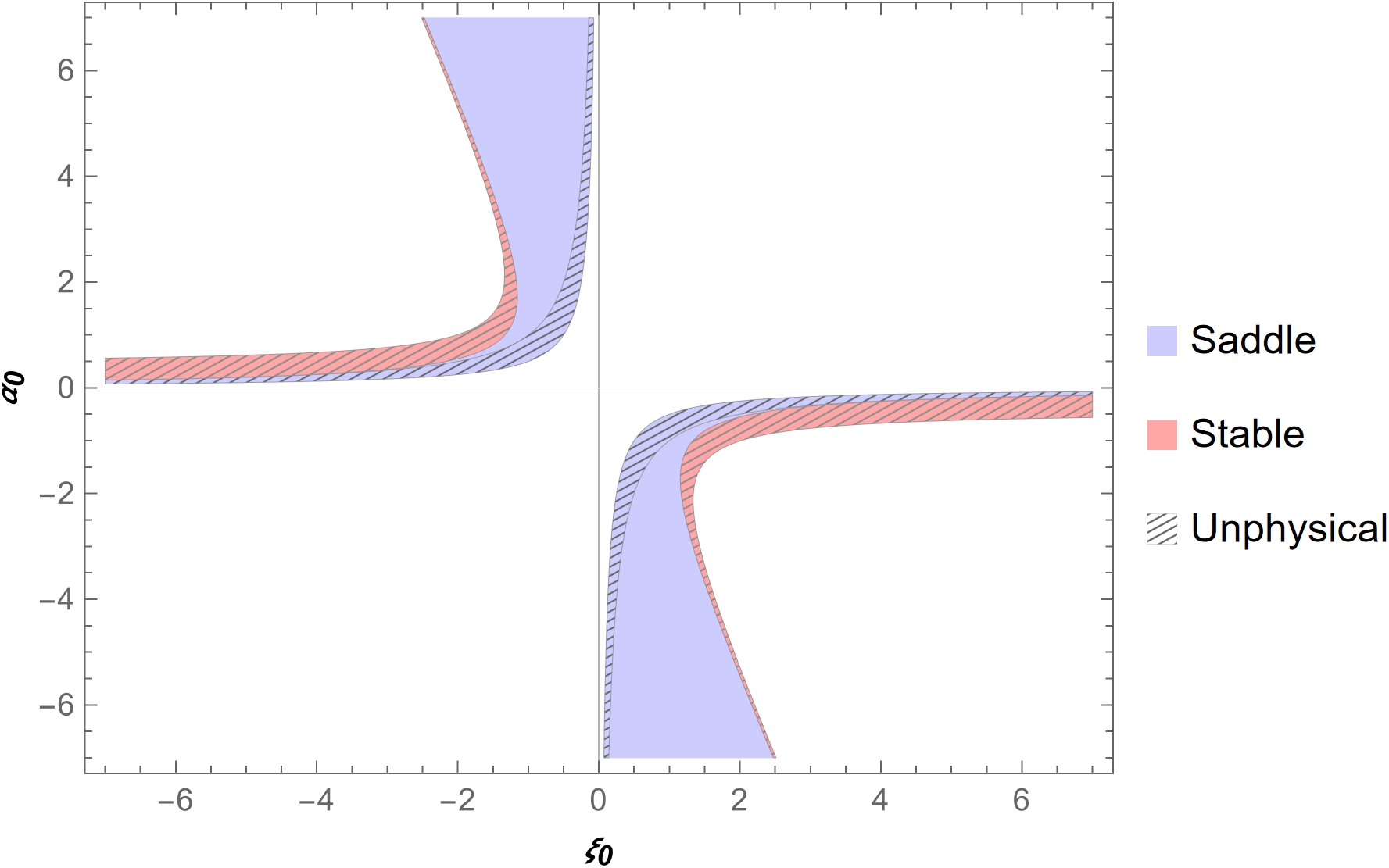}
    \caption{The parameter regions for the existence of critical point $A_{1}$.
        The shaded areas are considered unphysical due to either $\Omega_\phi>1$ (shaded pink area)
        or $\Omega_\phi<0$ (shaded blue area).}
    \label{xi0alpha}
\end{figure}
In Fig.\ref{xi0alpha},
the colored areas mark the possible regions of parameters for the critical point $A_1$ to exist.
The stable (pink) and saddle (blue) regions are divided by the line $\Omega_\phi=1$.
However, the conditions $0\le\Omega_m,\Omega_\phi\le1$ further rule out the unphysical regions (shaded) of the parameters $\xi_0$ and $\alpha_0$.

\textbullet\textbf{For critical point $A_{2}$},
the eigenvalues of the linearized perturbation matrix are
\begin{equation}
    \begin{split}
        \eta_{A_{2}}^{(1)}=&\frac{3}{2}+3\alpha_{0}\xi_{0} \,,\\
        \eta_{A_{2}}^{(2)}=&-\frac{3}{2}-9\xi_{0}^{2} \,.
    \end{split}
\end{equation}
\begin{figure}
    \centering
    \includegraphics[width=.9\linewidth]{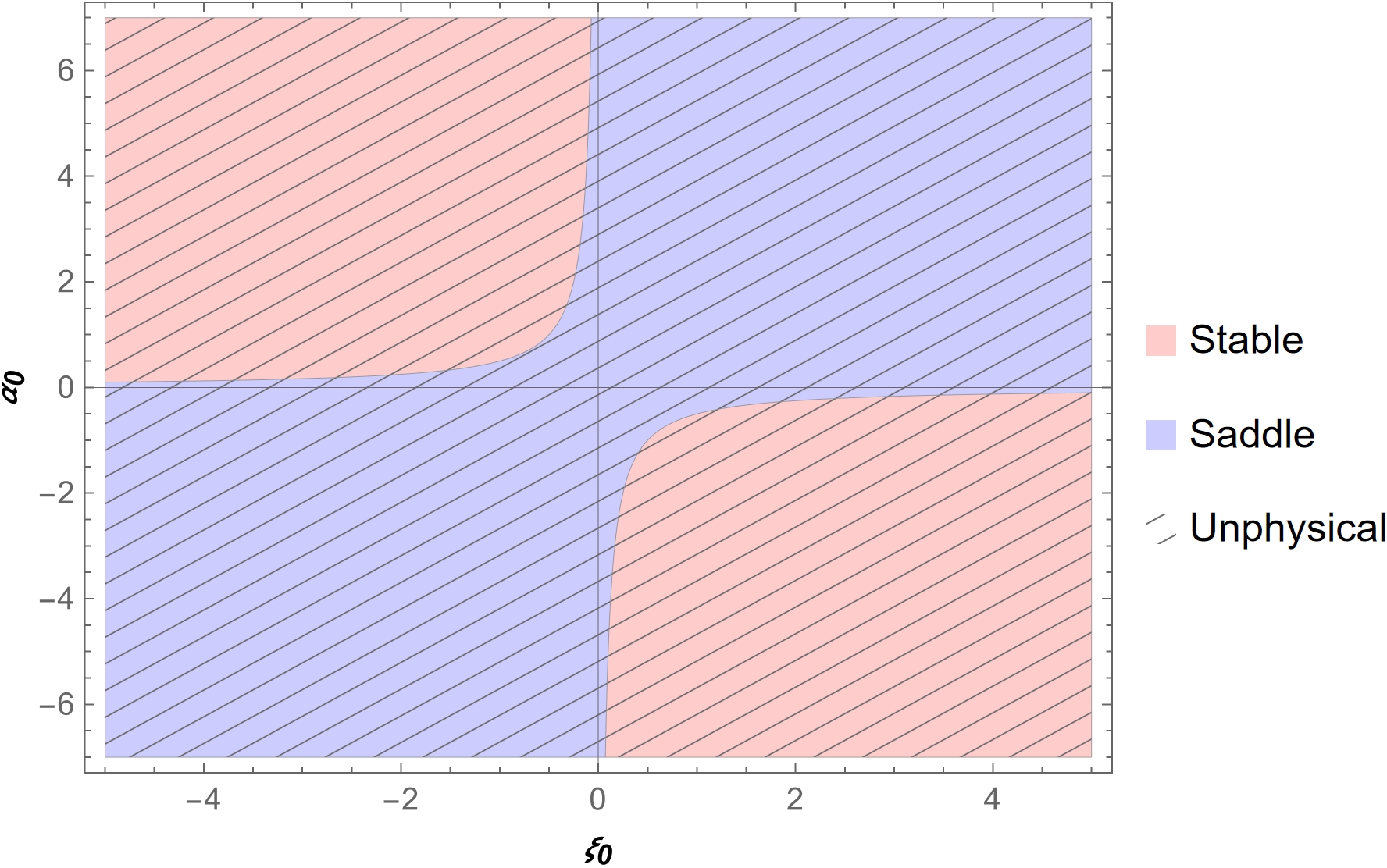}
    \caption{The parameter regions for the existence of critical point $A_{2}$.}
    \label{xi0alpha2}
\end{figure}
$A_2$ exists as long as $\alpha_{0} \ne 0$.
The sign of the first eigenvalue $\eta_{A_{2}}^{(1)}$ is bifurcated by the hyperbola $\alpha_{0} \xi_{0}=-1/2$,
while the second eigenvalue $\eta_{A_{2}}^{(2)}$ is always negative for any real value of parameter $\xi_{0}$.
Therefore, the stable (pink) and saddle (blue) regions of $A_{2}$ are shown as Fig.\ref{xi0alpha2}.
However, we note that $A_2$ only exists when $\Omega_\phi < 0$.
This scenario is also known as the perfect fluid supra-dominated era\cite{UrenaLopez2005},
which is beyond the scope of the current work and is considered unphysical.
We constrain the parameters $\alpha_0$ and $\xi_0$ as such that
$0\le\Omega_m,\Omega_\phi\le1$ and this point $A_2$ will not exist.

\textbullet\textbf{For critical point $A_{3}$},
the eigenvalues of the linearized perturbation matrix are
\begin{equation}
    \begin{split}
        \eta_{A_{3}}^{(1)}=&-3-\alpha_{0}(\alpha_{0}+3\xi_{0}) \,,\\
        \eta_{A_{3}}^{(2)}=&-\frac{1}{2}(6+(\alpha_{0}+3\xi_{0})^{2}) \,.
    \end{split}
\end{equation}
\begin{figure}
    \centering
    \includegraphics[width=1\linewidth]{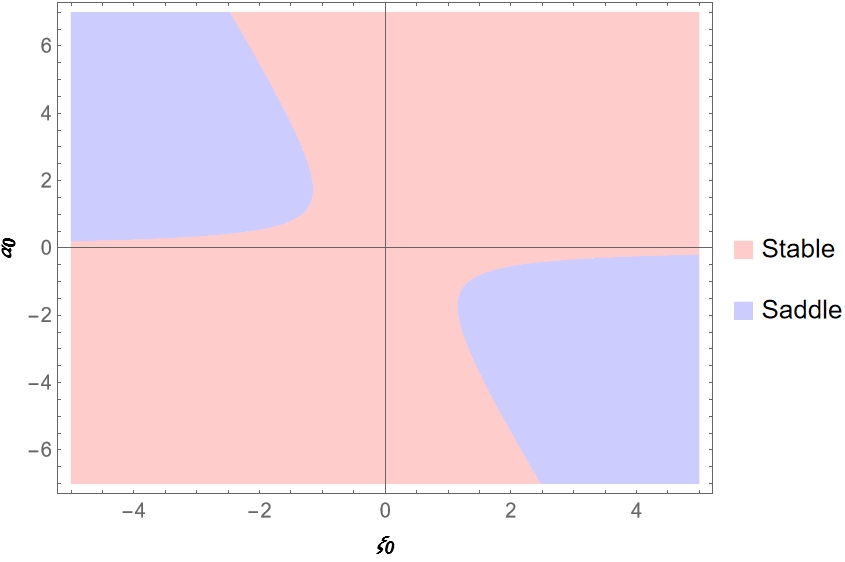}
    \caption{The parameter regions for the existence of critical point $A_{3}$.}
    \label{xi0alpha3}
\end{figure}
$A_3$ also exists for $\alpha_{0} \ne 0$ and any real $\xi_0$.
The second eigenvalue $\eta_{A_{3}}^{(2)}$ is always negative,
while the first one can have either sign.
The stable (pink) and saddle (blue) regions of $A_{3}$ are shown in Fig.\ref{xi0alpha3}.
We note that at this point the phantom field energy density parameter $\Omega_\phi = 1$,
which is invariant under the variation of the parameters $\alpha_{0}$ and $\xi_0$.
So, this point represents a phantom field dominated universe and is always physical.

The saddle and physically relevant regions of parameters for $A_1$ are subsets of the stable regions of $A_3$,
but only under the circumstance that $A_1$ exists
(i.e., the pair of parameters lies in the colored region depicted in Fig. \ref{xi0alpha}).
In this scenario,
the system will evolve from $A_1$ to $A_3$ in the cosmic history since $A_2$ does not exist for these parameters.
However, it is also possible that the parameters do not allow for the existence of $A_1$
(i.e., they fall outside the colored region depicted in Fig. \ref{xi0alpha}),
yet satisfy the physical conditions $0\le\Omega_m,\Omega_\phi\le1$.
Then, $A_3$ becomes the sole existing critical point
and the system will simply converge towards it.
The best-fit result of Model vP corresponds to the second scenario.

\begin{figure}[h]
    \centering
    \includegraphics[width=1\linewidth]{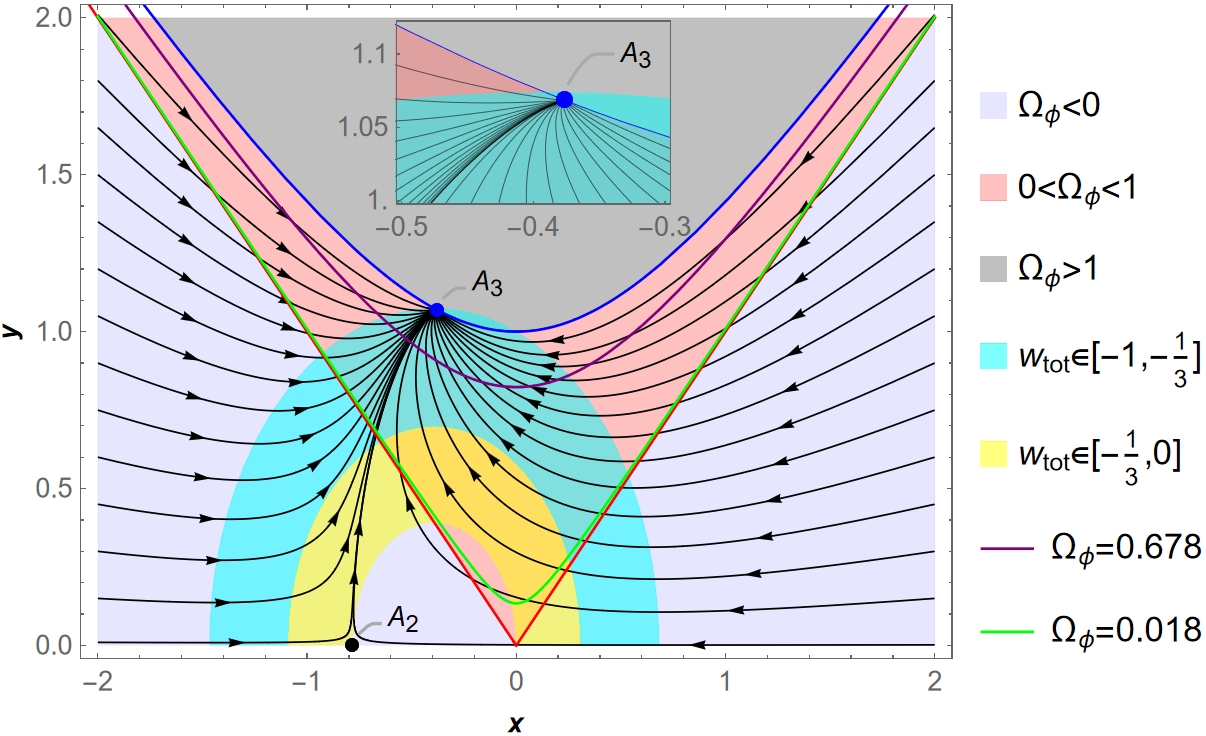}
    \caption{The phase space evolution of the dynamical system for Model vP.}
    \label{fig:xiangtu}
\end{figure}

Figure \ref{fig:xiangtu} shows the phase portrait of the viscous phantom field in the late universe.
The phase plane is divided into three regions:
(i) $\Omega_\phi> 1$ (the light gray region above the upper branch of the blue hyperbola $y^{2}-x^{2}=1$);
(ii) $0 < \Omega_\phi < 1$ (the region between the upper branch of the blue hyperbola $y^{2}-x^{2}=1$ and red straight lines $y=\pm x$);
and (iii) $\Omega_\phi < 0$ (the regions below the red straight lines).
The blue hyperbola $y^{2}-x^{2}=1$ represents all the possible attractors $A_3$ for different values of parameters $\{\alpha_0,\xi_0\}$.
The blue rounded marker indicates the specific attractor point $A_3$ for the best-fit parameters.
The purple hyperbola corresponds to $\Omega_{\phi0}=1-\Omega_{m0}=0.678$,
which is the best-fit density parameter of the phantom field at present time.
The green hyperbola corresponds to $\Omega_\phi=0.018$,
which is the value of $\Omega_\phi$ when $\Omega_m$ reaches its maximum in the cosmic history,
indicating the matter-dominated era calculated from the fitting result.

Moreover, the total EoS of the cosmic fluid can be expressed in terms of $x,y$ and $\zeta$ as
\begin{equation}\label{eqs:hubb}
    w_\text{tot} = \frac{p_\text{eff}+p_{m}}{\rho_{\phi}+\rho _{m} }=-x^2-y^2-\zeta\,.
\end{equation}
The universe is in accelerating phase if $-1< w_\text{tot}<-1/3$,
and decelerating if $-1/3< w_\text{tot}<0$.
When $w_\text{tot}<-1$, the universe will enter the super-acceleration and end in a Big Rig singularity.
For Model P which has no viscosity,
the region $-1\le w_\text{tot}=-x^2-y^2<-1/3$ intercepts with the blue hyperbola $y^2-x^2=1$ only at one point $(x,y)=(0,1)$,
which means that either the model needs severe fine-tuning or it will encounter big rip singularity.
This is independent of the fitting.
For Model vP, the regions $w_\text{tot}\in[-1,-1/3]$ and $w_\text{tot}\in[-1/3,0]$
calculated from the best-fit result are painted cyan and yellow in Fig. \ref{fig:xiangtu}, respectively.
One can see that the cyan region covers a portion of the blue hyperbola that represents all possibilities of $A_3$.
The best-fit $A_3$ is in this region.
Therefore, the presence of bulk viscosity can indeed allow the model evolve into an attractor that is still in the $w_\text{tot}>-1$ region,
and help avoid the cosmic singularity.

A heteroclinic orbit representing the evolutionary history of our universe
most likely starts somewhere near the green hyperbola (matter dominated era)
in the yellow region (decelerating phase with $-1/3\le w_\text{tot}\le0$),
crosses the purple hyperbola (present time) in the cyan region (accelerating phase),
and eventually reaches some point on the blue hyperbola (attractor $A_3$).
However, the preceding section of such a heteroclinic orbit seems to cross from the $\Omega_\phi<0$ region to $\Omega_\phi>0$ region in early time
and encounter a singularity of the phantom field EoS $w_\phi=p_\text{eff}/\rho_\phi$.
This issue arises because our analysis relies on late time observation data,
which do not capture the early stages of cosmic phase transitions.
Further research may be needed to address this typical concern rooting from the negative kinetic energy of phantom field.

\section{Statefinder diagnostic}
\label{statefinder}
The statefinder diagnostic uses the parameters $\{q,r,s\}$ that are derived from higher derivatives of the scale factor,
\begin{equation}
    q\equiv-\frac{\ddot a}{aH^2},\quad r\equiv\frac{\dddot{a}}{aH^3},\quad s\equiv\frac{r-1}{3(q-1/2)}\,,
\end{equation}
to differentiate various DE models.
For the current model, the parameters are
\begin{equation}
    \begin{split}
        q  =&\frac12 (1+3w_\phi \Omega_\phi),\\
        r  =&1-\frac{3}{2} \frac{{\diff} w_\phi}{{\diff} N} \Omega_\phi+\frac92w_\phi(1+w_\phi)\Omega_\phi,\\
        s  =&1- \frac{{\diff} w_\phi}{{\diff} N}\frac{1}{3w_\phi}+w_\phi.
    \end{split}
\end{equation}

\begin{figure}[h]
    \centering
    \includegraphics[width=0.8\linewidth]{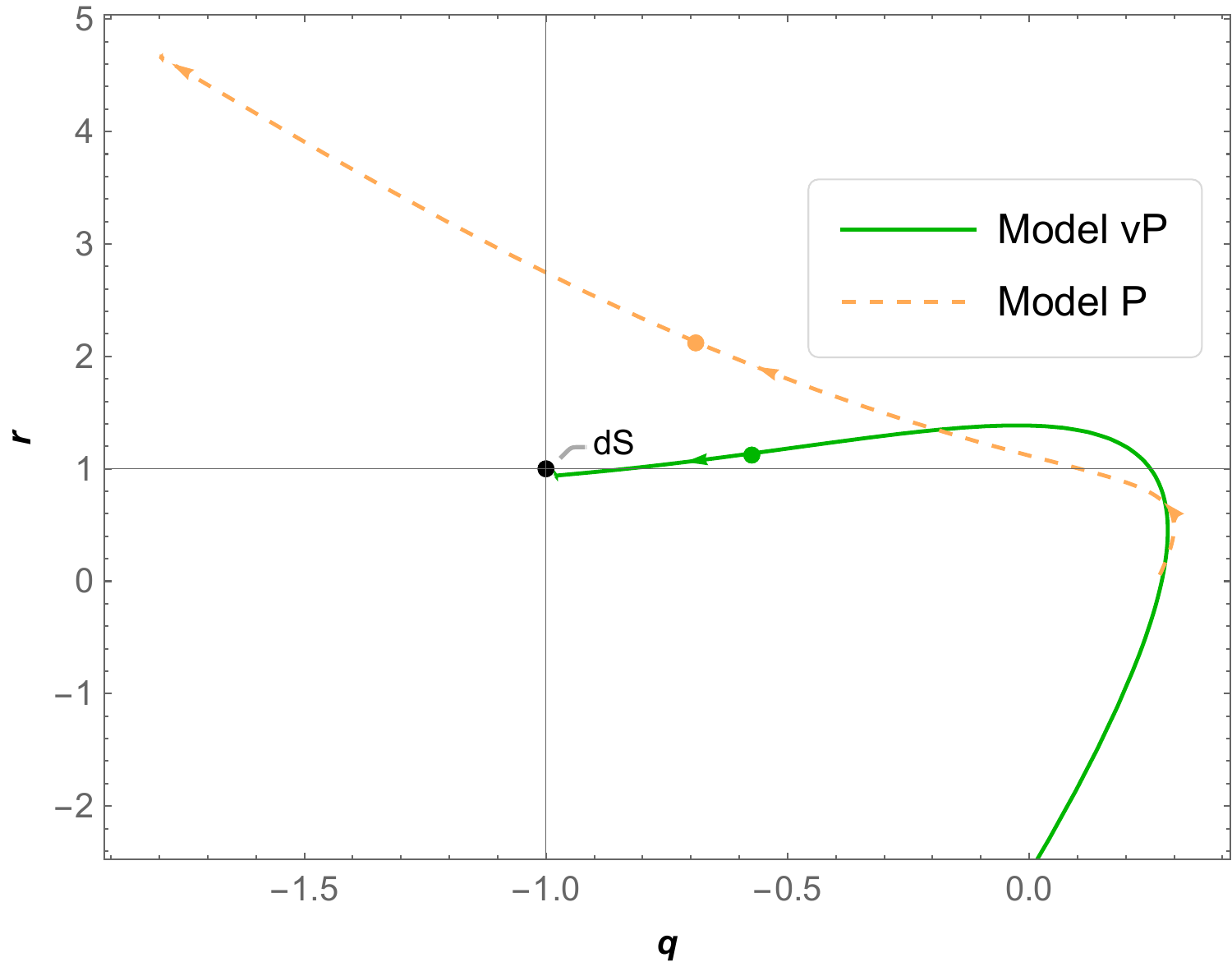}
    \caption{Evolving trajectories of the statefinder pairs in  the $q$-$r$ plane.
        The black dot $(-1,1)$ represents the de Sitter phase,
        and the solid dots on the lines represent the current state of the models.}
    \label{fig:rq}
\end{figure}

\begin{figure}[h]
    \centering
    \includegraphics[width=0.8\linewidth]{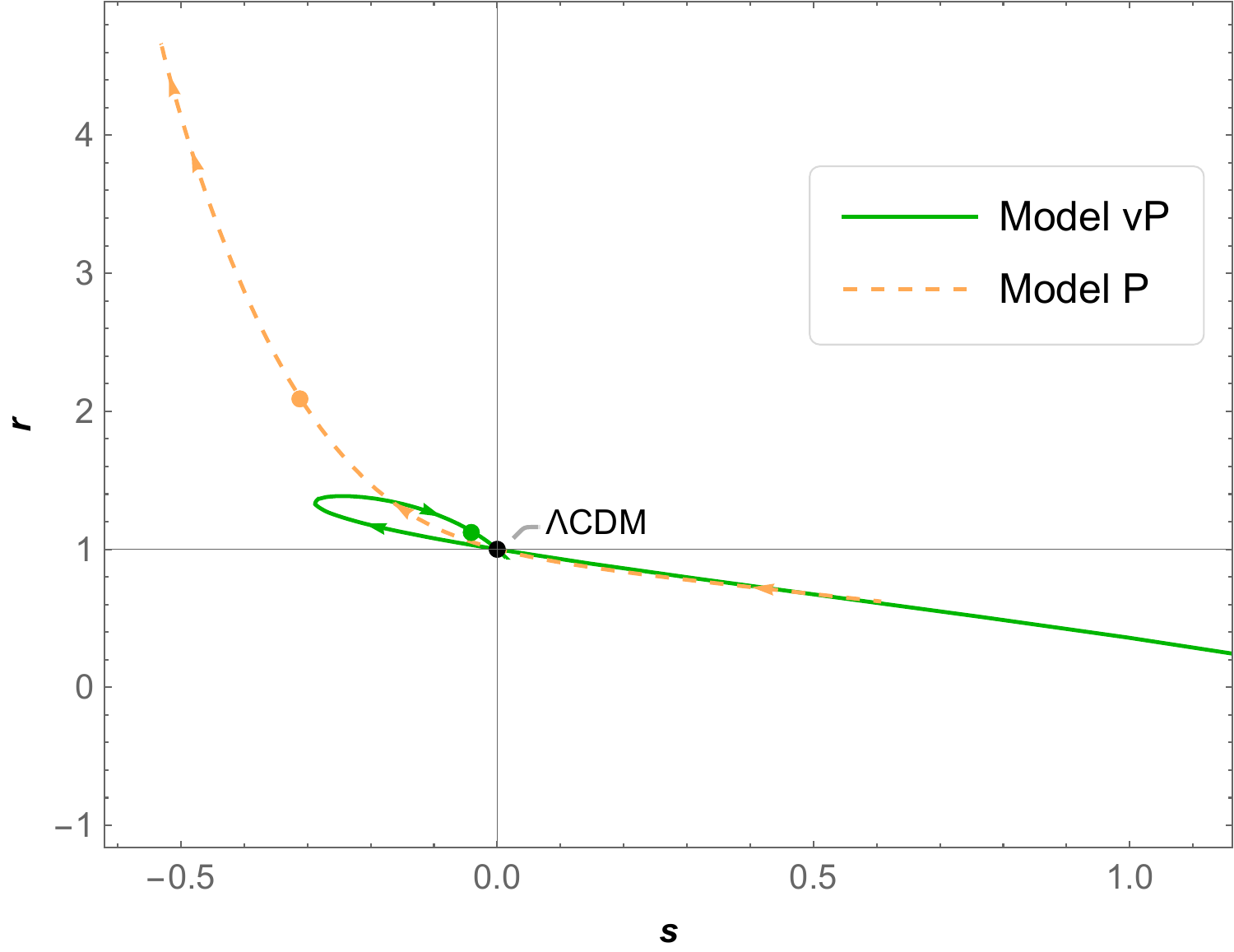}
    \caption{Evolving trajectories of the statefinder pairs in the $r$-$s$ plane.
        The black dot $(0,1)$ represents the $\Lambda $CDM model,
        and the solid dots on the lines represent the current state of the models.
    }
    \label{fig:rs}
\end{figure}

Using the best-fit parameters, we plot the $r$-$q$ and $r$-$s$ trajectories
for both Model P and Model vP in Figs. \ref{fig:rq} and \ref{fig:rs}, respectively.
The present time is indicated by the rounded marker on the curves.
As shown in Fig. \ref{fig:rq},
Model P exhibits a monotonically increasing deviation from the de Sitter point in the future
and asymptotes to a point in the region with $q<0,\:r>1$.
In contrast, Model vP will asymptotically approach a stable fixed point $(q, r) = (-0.9794, 0.9391)$ near the de Sitter point.
Fig. \ref{fig:rs} reveals that both models have $r < 1$ and $s > 0$ in the early universe,
and they both cross the $\Lambda$CDM point $(0, 1)$, but have different behaviors afterwards.
Model P moves away from the $\Lambda$CDM point,
while Model vP returns to it and passes it again before converging to a nearby fixed point $(s, r) = (0.0137, 0.9391)$ in the far future.

\begin{figure}[!h]
    \centering
    \includegraphics[width=0.8\linewidth]{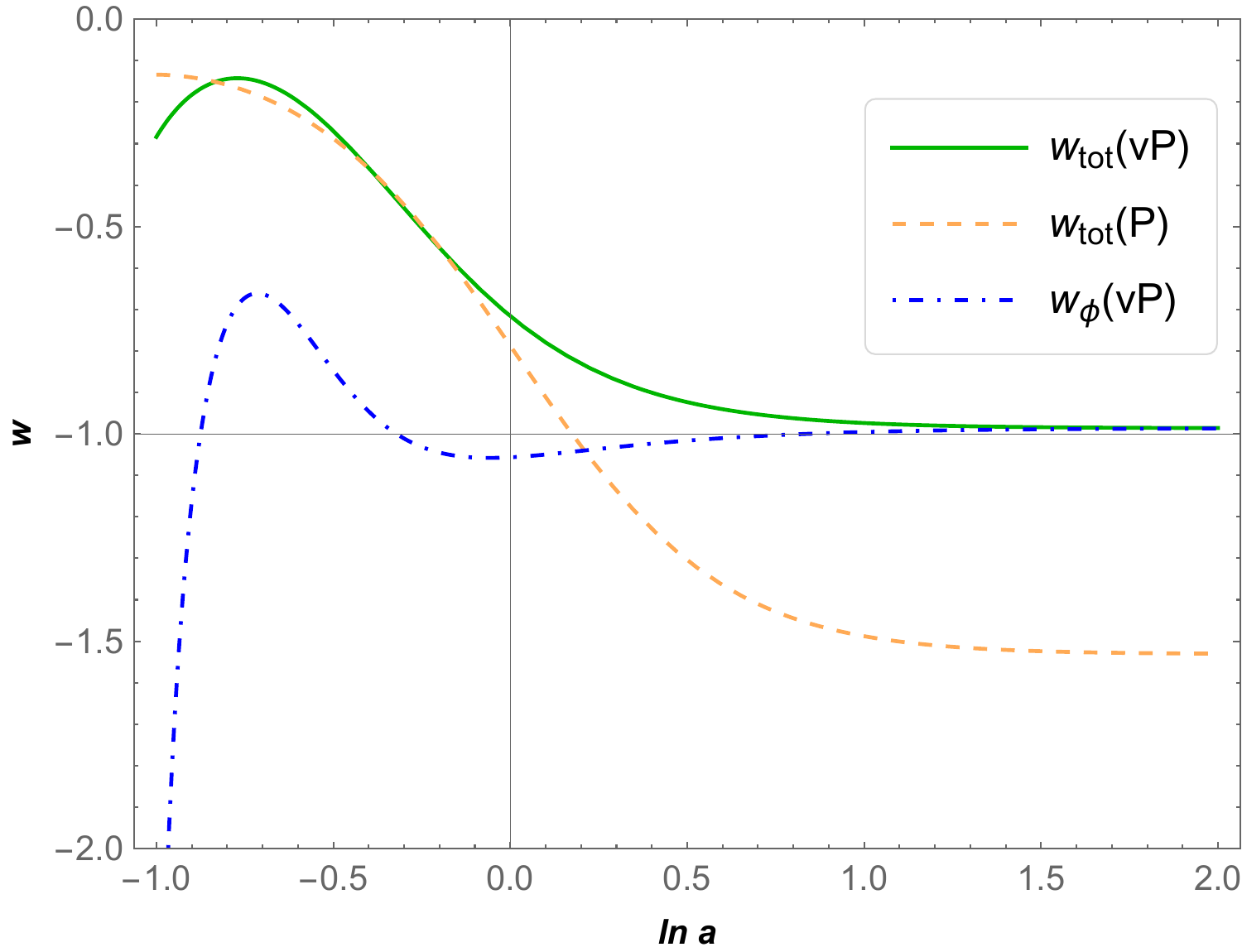}
    \caption{The evolution of the total EoSs, $w_\text{tot}$, of Model P and Model vP,
        as well as the effective EoS, $w_\phi$ of Model vP.}
    \label{fig:weff}
\end{figure}

The effective EoS parameters of the total cosmic fluid, $w_\text{tot}$, for both Model P and vP are plotted in Fig. \ref{fig:weff}.
As shown in the figure, $w_\text{tot}$ of Model vP remains above $-1$ and asymptotes to $-0.9863$ in the infinite future,
which implies that the universe will avoid the big rip singularity.
This result is consistent with the dynamical analysis in the previous section.
Moreover, the effective EoS parameter, $w_\phi$, of the viscous phantom field is also plotted.
One can see that it can cross the phantom divide during the cosmic evolution.
In this sense, the phantom field model with viscosity is an effective quintom model.
At present time, the effective EoSs of the viscous phantom field and the total cosmic fluid are $-1.0565$ and $-0.7159$, respectively.

\begin{figure}
    \centering
    \includegraphics[width=0.8\linewidth]{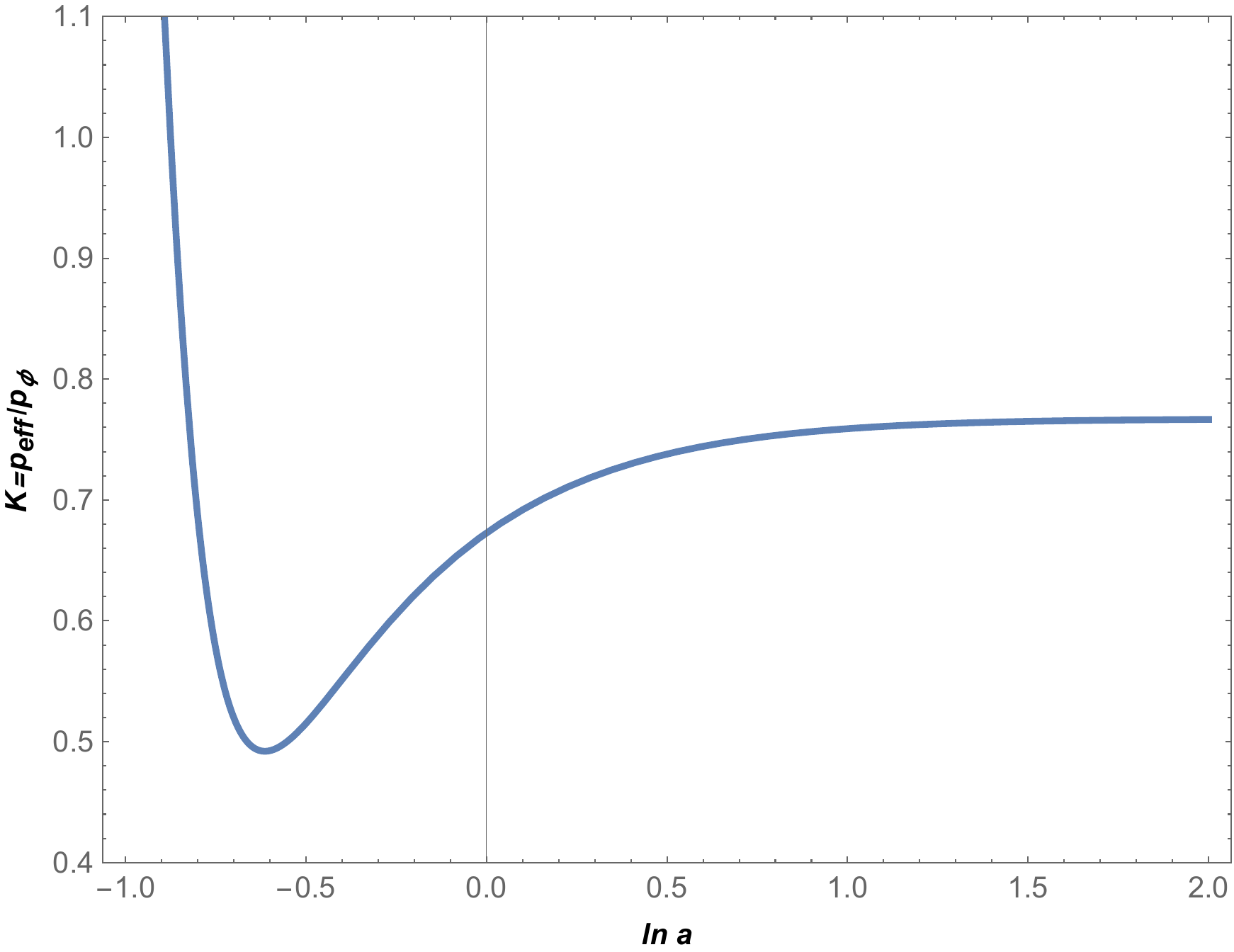}
    \caption{Evolution of the ratio of effective pressure $p_{eff}$ to scalar field pressure $p_{\phi}$.}
    \label{fig:peff}
\end{figure}

We also investigate the effect of viscosity on the evolutionary history
by computing the ratio $K$ of the effective pressure to the intrinsic pressure of the phantom field,
which is defined as
\begin{equation}
    K\equiv \frac{p_{eff}}{p_{\phi}}\sim 1+\dot{\phi}\rm{e}^{\frac{\alpha_{0}}{2}\phi}\,.
\end{equation}
Fig. \ref{fig:peff} shows the evolution of $K$ as a function of $\ln a$.
It is evident that viscosity has a significant impact on reducing the pressure of the phantom field,
especially around $z = 0.8475$ ($\ln a = -0.6138$),
where $K$ reaches a minimum value of about $0.4921$ and the pressure of the field is only a half of the intrinsic pressure.
At present ($z = 0$), the effective pressure is about one third ($32.73\%$) lower than the intrinsic pressure due to viscosity.
In the future, viscosity will continue to reduce the intrinsic pressure by about one quarter, as $K$ asymptotes to $0.7672$.

\section{Conclusion and discussions}
\label{conclude}
Recent cosmological observations suggest that the late-time acceleration of the universe is more likely driven by phantom DE rather than quintessence DE.
However, phantom DE models typically encounter the big rip singularity problem.
In this work, we explore the possibility of avoiding the big rip singularity by introducing bulk viscosity in the phantom field model.
We constrain the model parameters using the latest SNIa and H(z) data.
The results indicate that the data favor a significant bulk viscosity over a non-constant potential term for the phantom field.

We then perform a dynamical analysis of the viscous phantom field model using the best-fit values of the parameters.
We find that the only stable and physical attractor of the autonomous system is $A_3$, which corresponds to a phantom-dominated era.
The other critical points, may that be either unstable, unphysical, or cannot exist for the best-fit parameters,
do not represent different cosmological eras,
since the two variables of the autonomous system only correspond to the kinetic and potential energy.
We also plot the possible curves that represent the matter-dominated era and the present epoch on the phase portrait.
The heteroclinic orbit that describes our universe is expected to cross these curves and converge to the $A_3$ attractor.
However, we cannot trust the early part of the orbit or the critical points that precede the matter-dominated era,
since we only fit the model for the late-time behavior.
Due to bulk viscosity, some combinations of the parameters may allow the $A_3$ attractor have a total EoS $w_\text{tot}$ of cosmic fluid
that falls within the accelerating region $-1<w_\text{tot}<-1/3$.
This implies that the phantom universe may avoid the big rip singularity and end in a state with $w_\text{tot}>-1$ with the presence of bulk viscosity.

We apply the statefinder diagnostic to the viscous phantom field model and compare it with the non-viscous model.
We find that the statefinder parameters of the viscous model do not diverge monotonically from the $\Lambda$CDM point,
but rather approach a nearby fixed point asymptotically.
The bulk viscosity acts as a dissipative force that lowers the intrinsic pressure of the phantom field.
The viscosity has its maximum effect around $z = 0.8475$,
where it lowers the intrinsic pressure by about a half.
At the present time ($z = 0$), the viscosity reduces the intrinsic pressure by a third,
resulting in a lower effective pressure.
In the asymptotic future, the viscosity will still lower the intrinsic pressure by a quarter.
This reduction enables the universe to escape the big rip singularity.

For simplicity, we have limited our study to a specific form of the phantom field potential and the bulk viscosity,
which may not be the most general or realistic choice.
It would be worthwhile to explore other forms of potential and viscosity that can fit the observational data and avoid the big rip singularity.
Due to the scope of our work, we have only focused on the late-time behavior of the viscous phantom field model,
and ignored the early-time dynamics that may involve other mechanisms.
A more comprehensive study should include the full history of the universe and investigate the transitions between different cosmological eras.
We also use conventional methods of data analysis and dynamical analysis, which may have some limitations or biases.
Future research could employ artificial intelligence techniques to improve the accuracy and efficiency of data fitting,
parameter estimation, and model selection\cite{Cheng2018,Vagnozzi2018}.
\section*{Acknowledgement}
\label{ackn}
This work is supported by the National Science Foundation of China under Grant no. 12105179.
%\appendix

\bibliography{ref}
\end{document}